\documentclass[a4paper]{article}

\usepackage[english]{babel}
\usepackage[utf8]{inputenc}
\usepackage{amsmath}
\usepackage{graphicx}
\usepackage{epsfig}
\usepackage[colorinlistoftodos]{todonotes}
\usepackage[hidelinks]{hyperref}
\usepackage[margin=1.2in]{geometry}
\usepackage{natbib}
\usepackage{breqn}
\usepackage{bm}
\usepackage{bbm}
\usepackage{calc}

\usepackage{amsthm}
\usepackage{amsmath}               
  {
      \theoremstyle{plain}
      
  }

\usepackage{enumitem}

\title{Bayesian Causal Inference in Sequentially Randomized Experiments with Noncompliance}

\author{Jingying Zeng}

\date{December 2021}

\begin{document}
\maketitle

\section{Abstract}
Scientific researchers utilize randomized experiments to draw casual statements. Most early studies as well as current work on experiments with sequential intervention decisions has been focusing on estimating the causal effects among sequential treatments, ignoring the non-compliance issues that experimental units might not be compliant with the treatment assignments that they were originally allocated. A series of methodologies have been developed to address the non-compliance issues in randomized experiments with time-fixed treatment. However, to our best knowledge, there is little literature studies on the non-compliance issues in sequential experiments settings. In this paper, we 
go beyond the traditional methods using per-protocol, as-treated, or intention-to-treat analysis and propose a latent mixture Bayesian framework to estimate the sample-average treatment effect in sequential experiment having non-compliance concerns.

\section{Introduction}

Causal inference has been widely used in the fields of biology, psychology, and social science. In statistics literature, the most commonly used framework is the potential outcomes framework that was introduced by \citet{neyman1923application} for randomized experiments and further generalized by  \citet{rubin1974estimating, rubin1977assignment} for observational data. Under potential outcomes framework, much research on stratification, matching, and weighting to adjust for confounding issues has been focusing on static treatments at a single time point. In real-world, dynamic treatment strategies are also common in practice. For instance, in many clinical trails, cancer patients are firstly randomized to different chemotherapy regimens based on their baseline characteristics. At six-month follow-up, intensive treatments might no longer necessary for some patients while more aggressive treatments might be essential for the others. A new intervention is needed to dynamically adapt the change so that the patients are re-randomized again based on their prior treatments and their individual response to the disease. 

\begin{figure}[h]
    \centering
    \includegraphics[width=0.6\textwidth]{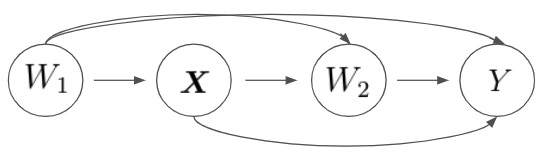}
    \caption{Two-Stage Sequentially Randomized Experiment}
    \label{fig:sequentially_randomized_experiment}
\end{figure}

The experiments with sequential intervention decisions at different time points are referred to as sequentially randomized experiments, and those subsequent outcomes are referred to as time-varying confounders \citep{shinohara2013estimating}. A two-stage sequentially randomized experiment with full compliance can be graphically represents as Figure \ref{fig:sequentially_randomized_experiment}, where $W_1$, $W_2$ are the treatments applied at the first and the second time points, $\bm{X}$ is the intermediate covariates measured, and $Y$ is the outcome of interest. Since the causal effects of the subsequent treatments are confounded by the intermediate outcomes, traditional approaches developed for time-fixed treatments can not be applied to this dynamic setting. \citet{robins1986new} broadened the potential outcome frameworks into a general theory of counterfactuals, which brought to light the causal analysis of sequentially randomized experiments that involve time-varying treatments and covariates.

In the last decade, there has been a growing interest in causal inference of dynamic strategies. In statistics literature, the most common estimand of sequential treatments is the comparison between treatments $(w_1,w_2)$ and $(w^'_1, w^'_2)$, namely sample-average treatment effect, defined as 
\begin{equation*}
    \tau_{w_1,w_2,w^'_1, w^'_2} = \frac{1}{N} \sum_{i=1}^{N} \Big[ Y_i(w_1,w_2) - Y_i(w^'_1, w^'_2) \Big].
\end{equation*}
From frequentest perspectives, the early methods were developed based on sequential ignorability assumption and g-formula proposed by \citet{robins1986new}. \citet{robins2000marginal} generalized the inverse probability weighting (IPW) estimator using stabilized IP weights, which is sensitive to the parametric model of treatment conditional on the confounders $\bm{X}$. The doubly robust estimator using marginal structural models (MSMs), proposed by \citet{bang2005doubly}, uses  "clever covariate"  in the sequential regression to attain double robustness property. \citet{keil2014parametric} proposed a parametric g-estimator in estimating causal effects. With g-formula, parametric modeling leads to the most efficient estimation on causal effects, but when models are misspecified, the estimation can be bias. \citet{zhou2020residual} introduced the idea of "residual balancing" in estimating the weights for MSMs in IPW estimator, which helps in reducing the bias resulting from model misspecification. On the other hand, Bayesian inference of sequentially randomized experiments approaches the problem from treating missing potential outcomes as unknown parameters so that missing potential outcomes can be simulated from the posterior predictive distributions of missing potential outcomes given the observed data \citep{rubin1978bayesian}. Based on this missing data perspective, \citet{zajonc2012bayesian} proposed a Bayesian framework in estimating the sample average treatment effects and finding the optimal treatment regimes. \citet{keil2018bayesian} developed a parametric Bayesian g-formula by using shrinkage priors and g-formula to combine the posterior draws.

Randomized experiments are reliable in drawing causal statements. However, noncompliance to assigned treatments is the biggest concern in analyzing randomized experiments since it breaks the original randomization in creating comparable groups. Noncompliance occurs when subjects do not take the treatments that they were initially allocated. Traditionally, in the analyses of randomized experiments, three populations are typically considered, namely per protocol population, as-treated population, and intention-to-treat (ITT) population. Per protocol analyses only include subjects who complied with their treatment assignments. As-treated analyses exclude the information about treatment assignments and only consider the actual treatment received. On the other hand, ITT analyses focus only on treatment assignment rather than treatment recipient, which preserves the benefits of randomization. None of these traditional methods are valid in estimating causal effects due to the following reasons: per protocol analyses ignored the fact that receiving treatment or not is self-selected so the per protocol population is not representative of the target population; as-treated population does not take into account the foundamental benefits of randomization; and the causal effects estimated by ITT population are the effects of treatment assignments on the outcome of interest rather than the causal effects of the receipt of treatment \citep{imbens2015causal}. 

Much research has been done on noncompliance issues for randomized experiment with interventions at a single time. Starting with \citet{angrist1995identification}, a series research on randomized experiments with noncompliance approaches the problem from the perspective of instrumental variables and potential outcomes to estimate the local average treatment effects (LATE), also referred to as Complier Average Treatment Effect (CATE) by \citet{imbens1997estimating}. Under the assumption of exclusion restriction; that is, for noncompliers, treatment assignment have no impact on the outcome of interest, \citet{angrist1996identification} proposed a non-parametric method in estimating LATE by the ratio of the ITT effect of primary outcome of interest and the ITT effect of treatment received. In recent years, research on Bayesian inference in analyzing noncompliance has become popular. Model-based Bayesian approaches based on posterior predictive distribution of missing potential outcomes have been discussed in several papers 
\citep{imbens1997bayesian, hirano2000assessing, rubin2010bayesian, imbens2015causal}.


Noncompliance issues also exist in sequentially randomized experiments. To our best of knowledge, there is little literature in investigating noncompliance issue in such sequential experiments. As oppose to the traditional methods using per protocol population, as-treated population, and intention-to-treat (ITT) population, this study proposed a latent mixture Bayesian framework for sequential experiments with non-compliance, and derived the posterior distributions for inference.

\section{Methodology}
When having time-varying treatments, the method that we proposed in Chapter 3 is not a simple extension to adjust time-varying confounders, since the conditional assumption that 
\begin{equation}\label{cond_not_hold}
    (W_1,W_2) \perp Y(w_1,w_2) \big| \bm{X}
\end{equation}
does not hold anymore. Such conditional assumption can be evaluated from the perspective of graph theory.

\begin{figure}[h]
\centering
\includegraphics[width=0.8\textwidth]{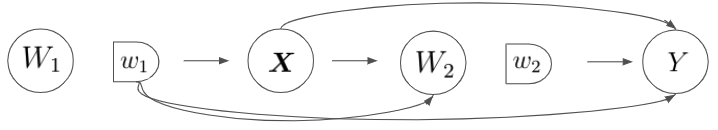}
\caption{Node Splitting in Creating SWIG of the Graph in Figure \ref{fig:sequentially_randomized_experiment} for the Hypothetical Intervention on $\{W_1, W_2\}$.}
    \label{sequential1}
\includegraphics[width=0.8\textwidth]{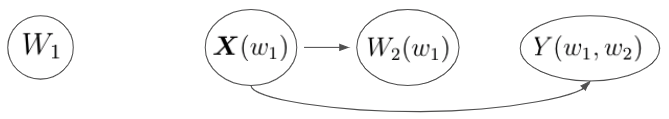}
\caption{SWIG of the Graph in Figure \ref{fig:sequentially_randomized_experiment} for the $W_1=w_1$ and $W_2=w_2$ intervention.}
    \label{sequential2}
\end{figure}

\begin{figure}[h]
    \centering
    \includegraphics[width=0.6\textwidth]{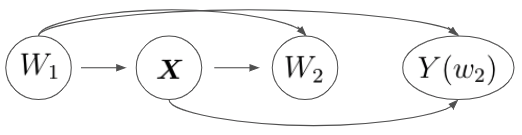}
    \caption{SWIG of the Graph in Figure \ref{fig:sequentially_randomized_experiment} for the $W_2=w_2$ intervention.}
    \label{fig:sequential3}
\end{figure}

For the hypothetical intervention of $W_1=w_1$ and $W_2=w_2$ made in the system, one can create a single-world intervention graph (SWIG) for the graphical model in Figure \ref{fig:sequentially_randomized_experiment} using node splitting (Figure \ref{sequential1}) and leaving out the deterministic components. As can be seen from the corresponding SWIG in Figure \ref{sequential2}, the conditional independence of $W_1 \perp Y(w_1,w_2)$ holds since $W_1$ and $Y(w_1,w_2)$ are in two different subgraphs in the SWIG. Additionally, in the path $Y(w_1,w_2) \leftarrow \bm{X}(w_1) \rightarrow W_2(w_1)$, if one conditions on $\bm{X}(w_1)$,  $W_2(w_1)$ and $Y(w_1,w_2)$ are d-separated, which implies the conditional independence $W_2(w_1) \perp  Y(w_1,w_2) \big| \bm{X}(w_1)$. Therefore, for sequentially randomized experiments, the assumption (\ref{cond_not_hold}) needs to be adjusted for time-varying variables in estimating causal effects. Similarly, the SWIG for the hypothetical intervention of $W_2=w_2$ can be constructed as Figure \ref{fig:sequential3}. $W_2$ and $Y(w_2)$ are d-separated by the set $\{ W_1,\bm{X} \}$, indicating the conditional dependence $W_2 \perp Y(w_2) \big| W_1,\bm{X}$.

\subsection{Statistical setting}
Here, we focus on two-stage sequentially randomized experiment, with the rationale discussed in the paper \citep{zajonc2012bayesian} that adding additional time periods is a simple extension, which only increases notational complexities without being too interesting theoretically. 

Suppose we have $i=1,..., N$ i.i.d subjects, observed at two time periods $t = 1, 2$. At each of the two period, each subject $i$ has a binary treatment assignment $Z_{it} \in \{0,1\}$, where $Z_{it} = 1$ indicate assigning to the treatment group while $Z_{it} = 0$ indicate assigning to the control. There are four sequences of treatment combination. $(1,1)$ denote assigning to treatment group in both periods, $(0, 0)$ denote assigning to control group in both periods, $(1,0)$ and $(0,1)$ denote assigning to treatment group only in the first and second period respectively. Treatment assignment at the first-period depends on the observed baseline covariates $\bm{X}_{i1}$, such as age, weight, race, or laboratory tests, measured prior to the treatment assignment. The treatment assignment at the second-period depends on the observed intermediate outcome $\bm{X}_{i2}^{obs}$, assessed after the first receipt of treatment but prior to the second assignment. We also observe the receipt of treatment, which is the treatment of primary interest, denoted as $W_{it}^{obs} \in \{0,1\}$. Since each subject might or might not choose to comply with his or her treatment assignment, noncompliance occurs when $Z_{it} \neq W_{it}^{obs}$. The compliance type is an instinct characteristic of an individual. Each subject's compliance behavior can be categorized into one of the following four compliance type $C_{it}$ at each stage $t$ \citep{imbens1997bayesian, hirano2000assessing}, defined as
\begin{equation*}
	C_{it}= \begin{cases}
	\text{Nevertakers (nt)} \\
	\text{Compliers (co)} \\
	\text{Defiers (df)} \\
	\text{Alwaystakers (at)} 
	\end{cases}. 
\end{equation*}

\begin{figure}[h!]
	\centering
	\includegraphics[width=0.8\linewidth]{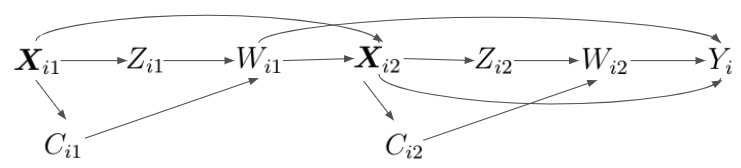}
	\caption{Two-Stage Sequentially Randomized Experiment with Noncompliance}
	\label{fig:sequentialwithnoncompliance}
\end{figure}

A final outcome of interest $Y_i^{obs}$, such as health score, is evaluated after the final treatment received. Figure \ref{fig:sequentialwithnoncompliance} illustrates this setup for a two-period sequentially randomized experiment with noncompliance. Under this graphical structure, we implicitly make an assumption that the effect of treatment assignment on the outcomes is only through the effect of the treatment received, which is a common assumption that researchers make when taking noncompliance into account.

In this chapter, we maintain the SUTVA assumption \citep{rubin1980randomization} to formulate the potential outcomes framework, that is, no different forms of the treatments and no interference among subjects. The outcome variables $W_{it}^{obs}$, $\bm{X}_{i2}^{obs}$, and $Y_i^{obs}$ are all associated with potential outcomes. $W_{it}(Z_{it})$ denotes the potential outcomes of the receipt of treatment, describing the treatment that would be received under hypothetical treatment assignment $Z_{it}$. As such, we link the potential outcomes of the treatment received to the treatment that is actually assign, that is,
$$W_{it}^{obs} = W_{it}(Z_{it}) = \begin{cases}
		W_{it}(0) & \text{if } Z_{it}=0\\
		W_{it}(1) & \text{if } Z_{it}=1
		\end{cases}.$$ 
Since only one of the potential outcomes of the treatment received can be possibly observed, we denote the missing one as $W_{it}^{mis}=W_{it}(1-Z_{it})$. Notice that, compliance type $C_{it}$ is a one-to-one function of $\Big( W_{i1}(0), W_{i1}(1), W_{i2}(0), W_{i2}(1) \Big)$ that can be written as
	\begin{equation*}
	C_{it}= \begin{cases}
	\text{Nevertakers (nt)} & \text{if } W_{it}(0)=0, W_{it}(1)=0\\
	\text{Compliers (co)} & \text{if } W_{it}(0)=0, W_{it}(1)=1\\
	\text{Defiers (df)} & \text{if } W_{it}(0)=1, W_{it}(1)=0\\
	\text{Alwaystakers (at)} & \text{if } W_{it}(0)=1, W_{it}(1)=1
	\end{cases}. 
	\end{equation*}

Similarly, $\bm{X}_{i2}\Big( W_{i1}(Z_{i1}), Z_{i1} \Big)$ is the potential outcomes of the intermediate measurements under hypothetical treatment assignment $Z_{i1}$ and the receipt of treatment $W_{i1}(Z_{i1})$. The observe intermediate outcome $\bm{X}_{i2}^{obs}$ equals
		\begin{align}
		\bm{X}_{i2}^{obs}
		=\begin{cases}
		\bm{X}_{i2}(0,0) & \text{if } W_{i1}^{obs}=0, Z_{i1}=0\\
		\bm{X}_{i2}(0,1) & \text{if } W_{i1}^{obs}=0, Z_{i1}=1\\
		\bm{X}_{i2}(1,0) & \text{if } W_{i1}^{obs}=1, Z_{i1}=0\\
		\bm{X}_{i2}(0,0) & \text{if } W_{i1}^{obs}=0, Z_{i1}=0
		\end{cases}, \nonumber
		\end{align} and the missing intermediate outcomes are $$\bm{X}_{i2}^{mis}=\Bigg( \bm{X}_{i2}\Big( 1-W_{i1}^{obs}, 0 \Big), \bm{X}_{i2}\Big( 1-W_{i1}^{obs}, 1 \Big) \Bigg).$$
Also, $Y_i\Big(Z_{i1},W_{i1}(Z_{i1}),Z_{i2},W_{i2}(Z_{i2})\Big)$ represents the potential outcomes of the final outcome of interest. We observe

$$Y_i^{obs} = Y_i\Big(Z_{i1}, W_{i1}^{obs}, Z_{i2}, W_{i2}^{obs}\Big)
		= Y_i \Big(Z_{i1}, W_{i1}(Z_{i1}), Z_{i2}, W_{i1}(Z_{i2}) \Big),$$
and the missing final outcomes are		
		\begin{align}
		\bm{Y}_i^{mis} =&\Bigg( Y_i\Big(0 , 1-W_{i1}^{obs}, 0, W_{i2}^{obs}\Big), Y_i\Big(0 , W_{i1}^{obs}, 0, 1-W_{i2}^{obs}\Big), Y_i\Big(0 , 1-W_{i1}^{obs}, 0, 1-W_{i2}^{obs}\Big),\nonumber\\
		& Y_i\Big(0 , 1-W_{i1}^{obs}, 1, W_{i2}^{obs}\Big), Y_i\Big(0 , W_{i1}^{obs}, 1, 1-W_{i2}^{obs}\Big), Y_i\Big(0 , 1-W_{i1}^{obs}, 1, 1-W_{i2}^{obs}\Big), \nonumber\\
		& Y_i\Big(1 , 1- W_{i1}^{obs}, 0, W_{i2}^{obs}\Big), Y_i\Big(1 , W_{i1}^{obs}, 0, 1-W_{i2}^{obs}\Big), Y_i\Big(1 , 1-W_{i1}^{obs}, 0, 1-W_{i2}^{obs}\Big), \nonumber\\
		&Y_i\Big(1, 1-W_{i1}^{obs}, 1, W_{i2}^{obs}\Big), Y_i\Big(1, W_{i1}^{obs}, 1, 1-W_{i2}^{obs}\Big), Y_i\Big(1, 1-W_{i1}^{obs}, 1, 1-W_{i2}^{obs}\Big)    \Bigg). \nonumber
		\end{align}

In order to identify the causal effects, one of the key assumptions for this framework is sequential unconfoundedness \citep{robins1986new}, that is, the observed treatments $\bm{W}_1^{obs}$ and $\bm{W}_2^{obs}$ are independent of the missing potential outcomes occurred in the past, and future potential outcomes conditioning on the past observed data. The second key assumption is the exclusion restriction assumptions for nevertakers and alwaystakers. Exclusion restriction \citep{imbens1997bayesian} assumes that the change in $Z_{it}$ will not affect the intermediate and the final outcomes for those subjects whose decisions in taking the treatment or not is uninfluenced by their treatment assignments. The exclusion restriction assumption holds by design in many applications, for instance, using a double blind study \citep{imbens2015causal}. The third assumption is the monotonicity assumption $W_{it}(1) \geq W_{it}(0)$ for $\forall i$ and $\forall t$, such that no defiers exists. Additionally, we assume the compliance types are time-invariant, i.e., $W_{i1}(0) = W_{i2}(0)$ and $W_{i1}(1) = W_{i2}(1)$, and assignment mechanisms do not depend on the compliance type.

Under exclusion restriction assumption, the potential outcome $Y_i(z_{i1},w_{i1},z_{i2},w_{i2})$ can be abbreviated as $Y(w_{i1},w_{i2})$ without ambiguity. Specifically, for compliers, \\
$Y_i(0,0) := Y_i\big( 0, W_{i1}(0), 0, W_{i2}(0) \big) = Y_i(0,0,0,0),
Y_i(0,1) := Y_i\big( 0, W_{i1}(0), 1, W_{i2}(1) \big) = Y_i(0,0,1,1) \\
Y_i(1,0) := Y_i\big( 1, W_{i1}(1), 0, W_{i2}(0) \big) = Y_i(1,1,0,0),
Y_i(1,1) := Y_i\big( 1, W_{i1}(1), 1, W_{i2}(1) \big) = Y_i(1,1,1,1).$\\
For alwaystakers, $Y_i(1,1) := Y_i\big( 1, W_{i1}(1), 1, W_{i2}(1) \big) 
 = Y_i\Big(0, W_i(0), 0, W_i(0)  \Big) 
 = Y_i\Big(0, W_i(0), 1, W_i(1)  \Big) \\
  = Y_i\Big(1, W_i(1), 0, W_i(0)  \Big) 
  =Y_i(1,1,1,1)$, but $Y_i(0,0)$, $Y_i(1,0)$, $Y_i(0,1)$ are not defined. For nevertakers,
$Y_i(0,0) := Y_i\big( 0, W_{i1}(0), 0, W_{i2}(0) \big) = Y_i\Big(0, W_i(0), 1, W_i(1)  \Big) = Y_i\Big(1, W_i(1), 0, W_i(0)  \Big) = Y_i\Big(1, W_i(1), 1, W_i(1)  \Big) =Y_i(0,0,0,0),$ but $Y_i(1,1)$, $Y_i(1,0)$, $Y_i(0,1)$ are not defined.

The target estimand, also known as the Local Average Treatment Effect (LATE) in literature, is the average of the subject-level comparisons between $Y_i(w_1,w_2)$ and $Y_i(w_1^{'},w_2^{'})$ among the subgroup of compliers, defined as
\begin{equation}\label{LATE}
\tau_{LATE} = \frac{1}{N_{co}} \sum_{i: C_i=co} \Big[  Y_i(w_1,w_2) - Y_i(w_1^{'},w_2^{'}) \Big],
\end{equation}
where $N_{co}$ is the total number of compliers. Statistically, the reason we may want to condition on complier is that the treatment effects would then be unconfounded. Practically, compliers are the group for whom intervention is actually possible, and thus understanding the treatment effect for such group would provide useful information in guiding interventions. This also explains another question: whether compliance status should be time-varying in a sequential experiment. The argument is that if compliance status were time-varying, it would introduce fundamental conceptual controversies. On the one hand, the target estimand is no longer valid since it is no longer clear what it means to condition on being a complier. Even more fundamentally, if compliance behavior could change during the course of an experiment when the experiment is sequential, it then means that non-compliance may no longer be an issue. For example, ``nevertaker'' is no longer ``never''-taker but rather ``temporarily'' non-taker, so that the researcher can simply wait for compliance status itself to change  so as to make the non-compliance issue go away. On the other hand, the non-compliance issue builds upon the assumption that compliance type is not easily intervened or altered.


\subsection{The Bayesian Perspective}
Let $\bm{Y}(1,1)$ denote the N-dimensional vector of the final potential outcomes under receipt of the treatment sequence $(1, 1)$, with the $i^{th}$ entry equals $Y_i(0,0)$. $\bm{Y}(0,1), \bm{Y}(1,0), \bm{Y}(1,1)$ are defined analogously. Similarly, $\bm{X}_2(1,1), \bm{X}_2(1,0), \bm{X}_2(0,1)$, and $\bm{X}_2(0,0)$ denote the N-dimensional vector of the intermediate potential outcomes; $\bm{W}_t(1)$ and $\bm{W}_t(0)$ denote the N-dimensional vector of the potential outcomes of treatment received at the $t^{th}$ period, $t=1,2$. 

Let $\bm{Y} = \Big( \bm{Y}(1,1),\bm{Y}(0,1), \bm{Y}(1,0), \bm{Y}(1,1) \Big)$, $\bm{X}_2 = \Big( \bm{X}_2(1,1), \bm{X}_2(1,0), \bm{X}_2(0,1), \bm{X}_2(0,0) \Big)$, and $\bm{W}_t = \Big( \bm{W}(1), \bm{W}(0) \Big)$, $t=1,2$. Also, let $\bm{C}$, $\bm{X}_1$, $\bm{Z}_1$, and $\bm{Z}_2$ be the N-dimensional vector of compliance type, baseline covariates, and treatment assignment at the first and the second period, respectively, where the $i^{th}$ entry of the vector equals the corresponding value of the $i^{th}$ subject. Causal estimand of the form $\tau \big( \bm{Y}, \bm{X}_2, \bm{X}_1, \bm{Z}_1 , \bm{Z}_2, \bm{W} \big)$ can be represented as a function of the observed and missing variables, 
\begin{equation*}
    \tau(\bm{Y}^{obs}, \bm{Y}^{mis}, \bm{X}_2^{obs}, \bm{X}_2^{mis}, \bm{X}_1, \bm{Z}_1, \bm{Z}_2,\bm{W}_1^{obs}, \bm{W}_1^{mis},\bm{W}_2^{obs}, \bm{W}_2^{mis})
\end{equation*}

From Bayesian Perspective, causal inference under potential outcomes framework is fundamentally a missing data problem, where the unobserved potential outcomes can be considered as unknown parameters. Finite-sample Bayesian inference is based on imputing missing potential outcomes given observed data so as to obtain the posterior distribution of the causal estimand \citep{rubin1978bayesian}. The quantity of interest is the posterior predictive distribution of the missing data based on observed data
\begin{align}
&P\big( \bm{Y}^{mis}, \bm{X}_2^{mis}, \bm{C} \, \big| \, \bm{Y}^{obs}, \bm{X}_2^{obs}, \bm{X}_1, \bm{Z}_1,     \bm{Z}_2 , \bm{W}_1^{obs},\bm{W}_2^{obs}  \big) \nonumber\\
=  &\int  P\big( \bm{Y}^{mis}, \bm{X}_2^{mis}, \bm{C}, \theta \, \big| \, \bm{Y}^{obs}, \bm{X}_2^{obs}, \bm{X}_1, \bm{Z}_1,     \bm{Z}_2 , \bm{W}_1^{obs},\bm{W}_2^{obs}  \big)  d \theta. \nonumber
\end{align}
One can further factorize as follows:
\begin{align}
& P\big( \bm{Y}^{mis}, \bm{X}_2^{mis}, \bm{C}, \theta \, \big| \, \bm{Y}^{obs}, \bm{X}_2^{obs}, \bm{X}_1, \bm{Z}_1,     \bm{Z}_2 , \bm{W}_1^{obs},\bm{W}_2^{obs}  \big) \nonumber \\
= & P(\theta | \bm{Y}^{obs}, \bm{X}_2^{obs}, \bm{X}_1, \bm{Z}_1, \bm{Z}_2, \bm{W}_1^{obs}, \bm{W}_2^{obs}) \label{4.3}\\
 \cdot & P(\bm{C} | \theta, \bm{Y}^{obs}, \bm{X}_2^{obs}, \bm{X}_1, \bm{Z}_1, \bm{Z}_2, \bm{W}_1, \bm{W}_2) \label{4.4}\\
 \cdot & P(\bm{Y}^{mis}, \bm{X}_2^{mis} | \bm{C}, \theta, \bm{Y}^{obs}, \bm{X}_2^{obs}, \bm{X}_1, \bm{Z}_1, \bm{Z}_2, \bm{W}_1^{obs}, \bm{W}_2^{obs}) \label{4.5}
\end{align}
so at a high-level, the sampling algorithm is:
\begin{itemize}
	\item Step 1: Sample $\theta \sim P(\theta | \bm{Y}^{obs}, \bm{X}_2^{obs}, \bm{X}_1, \bm{Z}_1, \bm{Z}_2, \bm{W}_1^{obs}, \bm{W}_2^{obs})$
	\item Step 2: Sample $\bm{C} \sim P(\bm{C} | \theta, \bm{Y}^{obs}, \bm{X}_2^{obs}, \bm{X}_1, \bm{Z}_1, \bm{Z}_2, \bm{W}_1, \bm{W}_2)$
	\item Step 3: Sample $\bm{Y}^{mis}, \bm{X}_2^{mis} \sim P(\bm{Y}^{mis}, \bm{X}_2^{mis} | \bm{C}, \theta, \bm{Y}^{obs}, \bm{X}_2^{obs}, \bm{X}_1, \bm{Z}_1, \bm{Z}_2, \bm{W}_1^{obs}, \bm{W}_2^{obs})$
\end{itemize}
Next, we are going to discuss each step in more details. For the first step,  notice that (\ref{4.3}) equals
\begin{equation}\label{firststep}
P(\theta | \bm{Y}^{obs}, \bm{X}_2^{obs}, \bm{X}_1, \bm{Z}_1, \bm{Z}_2, \bm{W}_1^{obs}, \bm{W}_2^{obs}) = \sum_{C_1',...,C_N'} P(\theta, \bm{C}' | \bm{Y}^{obs}, \bm{X}_2^{obs}, \bm{X}_1, \bm{Z}_1, \bm{Z}_2, \bm{W}_1^{obs}, \bm{W}_2^{obs}),
\end{equation}

Note that $P(\bm{X}_2^{obs} | \theta, \bm{C}', \bm{X}_1, \bm{Z}_1, \bm{Z}_2, \bm{W}_1^{obs}, \bm{W}_2^{obs}) = P(\bm{X}_2^{obs} | \theta, \bm{C}', \bm{X}_1, \bm{Z}_1,  \bm{W}_1^{obs})$ since the past would not depend on the future. Then under the assumption of conditional exchangeability and SUTVA, 

\begin{align}
P(\theta, \bm{C}' | \bm{Y}^{obs}, \bm{X}_2^{obs}, \bm{X}_1, \bm{Z}_1, \bm{Z}_2, \bm{W}_1, \bm{W}_2) &\propto P(\theta, \bm{C}', \bm{Y}^{obs}, \bm{X}_2^{obs} | \bm{X}_1, \bm{Z}_1, \bm{Z}_2, \bm{W}_1^{obs}, \bm{W}_2^{obs})  \nonumber \\
& = P(\bm{Y}^{obs} | \bm{X}_2^{obs}, \theta, \bm{C}', \bm{X}_1, \bm{Z}_1, \bm{Z}_2, \bm{W}_1^{obs}, \bm{W}_2^{obs}) \nonumber \\
& \quad  \cdot P(\bm{X}_2^{obs} | \theta, \bm{C}', \bm{X}_1, \bm{Z}_1, \bm{Z}_2, \bm{W}_1^{obs}, \bm{W}_2^{obs}) \nonumber\\
& \quad \cdot P(\theta, \bm{C}' | \bm{X}_1, \bm{Z}_1, \bm{Z}_2, \bm{W}_1^{obs}, \bm{W}_2^{obs}) \nonumber\\
& = \prod_{i=1}^{N} P(Y_i^{obs} | X_{i2}^{obs}, \theta, C_i', \bm{X}_{i1}, Z_{i1}, Z_{i2}, W_{i1}^{obs}, W_{i2}^{obs}) \nonumber \\
& \quad  \cdot \prod_{i=1}^{N}  P(X_{i2}^{obs} | \theta, C_i', \bm{X}_{i1}, Z_{i1},  W_{i1}^{obs}) \nonumber \\
& \quad \cdot P(\theta, \bm{C}' | \bm{X}_1, \bm{Z}_1, \bm{Z}_2, \bm{W}_1^{obs}, \bm{W}_2^{obs}). \nonumber
\end{align}
Furthermore, 
	\begin{align*}
	P(\theta, \bm{C}' | \bm{X}_1, \bm{Z}_1, \bm{Z}_2, \bm{W}_1^{obs}, \bm{W}_2^{obs}) &\propto P(\theta, \bm{C}', \bm{X}_1, \bm{Z}_1, \bm{Z}_2, \bm{W}_1^{obs}, \bm{W}_2^{obs}) \\
	&= P(\theta) \cdot P(\bm{C}' | \theta, \bm{X}_1) \cdot P(\bm{Z}_1, \bm{Z}_2, \bm{W}_1^{obs}, \bm{W}_2^{obs} | \bm{C}', \theta, \bm{X}_1),
	\end{align*}
where $P(\theta)$ is the prior, $P(\bm{C}'|\theta, \bm{X}_1)$ models how compliance type would depend on initial covariates and $\theta$, and for $P(\bm{Z}_1, \bm{Z}_2, \bm{W}_1^{obs}, \bm{W}_2^{obs} | \bm{C}', \theta, \bm{X}_1)$, we have
	\begin{align*}
	& P(\bm{Z}_1, \bm{Z}_2, \bm{W}_1^{obs}, \bm{W}_2^{obs} | \bm{C}', \theta, \bm{X}_1)\\
	&= P(\bm{W}_1^{obs}, \bm{W}_2^{obs} | \bm{Z}_1, \bm{Z}_2, \bm{C}', \theta, \bm{X}_1) \cdot P(\bm{Z}_1, \bm{Z}_2 | \bm{C}', \theta, \bm{X}_1) \\
	&= P(\bm{W}_2^{obs} | \bm{W}_1^{obs}, \bm{Z}_1, \bm{Z}_2, \bm{C}', \theta, \bm{X}_1) \cdot P(\bm{W}_1^{obs} | \bm{Z}_1, \bm{Z}_2, \bm{C}', \theta, \bm{X}_1) \cdot P(\bm{Z}_1, \bm{Z}_2 | \bm{C}', \theta, \bm{X}_1) \\
	&= P(\bm{W}_2^{obs} | \bm{Z}_2, \bm{C}') \cdot P(\bm{W}_1^{obs} | \bm{Z}_1, \bm{C}') \cdot P(\bm{Z}_1, \bm{Z}_2 | \bm{C}', \theta, \bm{X}_1) \\
	&= P(\bm{W}_2^{obs} | \bm{Z}_2, \bm{C}') \cdot P(\bm{W}_1^{obs} | \bm{Z}_1, \bm{C}') \cdot P(\bm{Z}_1, \bm{Z}_2 | \bm{X}_1)
	\end{align*}
The reason why $P(\bm{Z}_1, \bm{Z}_2 | \bm{C}', \theta, \bm{X}_1) = P(\bm{Z}_1, \bm{Z}_2 | \bm{X}_1)$ is because of the assumption that assignment mechanisms do not depend on compliance type. Therefore,
\begin{align}\label{ctheta}
      P(\theta, \bm{C}' | \bm{X}_1, \bm{Z}_1, \bm{Z}_2, \bm{W}_1^{obs}, \bm{W}_2^{obs}) &  \propto
	P(\theta) \cdot P(\bm{C}' | \theta, \bm{X}_1) \cdot  P(\bm{W}_2^{obs} | \bm{Z}_2, \bm{C}') \cdot P(\bm{W}_1^{obs} | Z_1, C_i') \nonumber \\
	&= P(\theta) \cdot \prod_{i=1}^{N} \Bigg[ P(C_i' | \theta, \bm{X}_{i1}) \cdot  P(W_{i2}^{obs} | Z_{i2}, C_i') \cdot P(W_{i1}^{obs} | Z_{i1}, C_i') \Bigg]
\end{align}
As can be seen from the derivation, assignment mechanism $P(\bm{Z}_1, \bm{Z}_2 | \bm{X}_1)$ can be dropped off from the posterior. Therefore, (\ref{firststep}) can be written as

\begin{align*}
    P(\theta | \bm{Y}^{obs}, \bm{X}_2^{obs}, \bm{X}_1, \bm{Z}_1, \bm{Z}_2, \bm{W}_1, \bm{W}_2)
\propto P(\theta) \cdot \prod_{i=1}^{N}  \sum_{C_i'} &  \cdot  \Bigg[ P(Y_i^{obs} | X_{i2}^{obs}, \theta, C_i', \bm{X}_{i1}, Z_{i1}, Z_{i2}, W_{i1}^{obs}, W_{i2}^{obs}) \nonumber \\
& \quad  \cdot   P(X_{i2}^{obs} | \theta, C_i', \bm{X}_{i1}, Z_{i1},  W_{i1}^{obs}) \nonumber \\
&  \cdot  P(C_i' | \theta, \bm{X}_{i1}) \cdot  P(W_{i2}^{obs} | Z_{i2}, C_i') \cdot P(W_{i1}^{obs} | Z_{i1}, C_i') \Bigg]
\end{align*}

For the second step, (\ref{4.4}) equals
\begin{align*}
P(\bm{C} | \theta, \bm{Y}^{obs}, \bm{X}_2^{obs}, \bm{X}_1, \bm{Z}_1, \bm{Z}_2, \bm{W}_1^{obs}, \bm{W}_2^{obs}) &= \frac{P(\bm{C}, \bm{Y}^{obs}, \bm{X}_2^{obs} | \theta, \bm{X}_1, \bm{Z}_1, \bm{Z}_2, \bm{W}_1^{obs}, \bm{W}_2^{obs})}{P(\bm{Y}^{obs}, \bm{X}_2^{obs} | \theta, \bm{X}_1, \bm{Z}_1, \bm{Z}_2, \bm{W}_1^{obs}, \bm{W}_2^{obs})} \\
&= \frac{P(\bm{C}, \bm{Y}^{obs}, \bm{X}_2^{obs} | \theta, \bm{X}_1, \bm{Z}_1, \bm{Z}_2, \bm{W}_1^{obs}, \bm{W}_2^{obs})}{ \sum_{C_1',...,C_N'} P(\bm{Y}^{obs}, \bm{X}_2^{obs}, \bm{C}' | \theta, \bm{X}_1, \bm{Z}_1, \bm{Z}_2, \bm{W}_1^{obs}, \bm{W}_2^{obs})},
\end{align*}
and
\begin{align}
 P(\bm{C}, \bm{Y}^{obs}, \bm{X}_2^{obs} | \theta, \bm{X}_1, \bm{Z}_1, \bm{Z}_2, \bm{W}_1^{obs}, \bm{W}_2^{obs}) 
& = P(\bm{Y}^{obs} | \bm{X}_2^{obs}, \theta, \bm{C}, \bm{X}_1, \bm{Z}_1, \bm{Z}_2, \bm{W}_1^{obs}, \bm{W}_2^{obs}) \nonumber \\
& \quad  \cdot P(\bm{X}_2^{obs} | \theta, \bm{C}, \bm{X}_1, \bm{Z}_1, \bm{Z}_2, \bm{W}_1^{obs}, \bm{W}_2^{obs}) \nonumber\\
& \quad \cdot P(\bm{C} | \theta, \bm{X}_1, \bm{Z}_1, \bm{Z}_2, \bm{W}_1^{obs}, \bm{W}_2^{obs}) \nonumber \\
& = \prod_{i=1}^{N} P(Y_i^{obs} | X_{i2}^{obs}, \theta, C_i', \bm{X}_{i1}, Z_{i1}, Z_{i2}, W_{i1}^{obs}, W_{i2}^{obs}) \nonumber \\
& \quad  \cdot \prod_{i=1}^{N}  P(X_{i2}^{obs} | \theta, C_i', \bm{X}_{i1}, Z_{i1},  W_{i1}^{obs}) \nonumber \\
& \quad \cdot P(\bm{C} | \theta, \bm{X}_1, \bm{Z}_1, \bm{Z}_2, \bm{W}_1^{obs}, \bm{W}_2^{obs}) \nonumber. 
\end{align}
Since
\begin{align*}
P(\bm{C}| \theta, \bm{X}_1, \bm{Z}_1, \bm{Z}_2, \bm{W}_1^{obs}, \bm{W}_2^{obs}) &\propto P(\bm{C}, \theta| \bm{X}_1, \bm{Z}_1, \bm{Z}_2, \bm{W}_1^{obs}, \bm{W}_2^{obs}) \\
&\propto P(\bm{W}_2^{obs} | \bm{Z}_2, \bm{C}) \cdot P(\bm{W}_1^{obs} | \bm{Z}_1, \bm{C}) \cdot P(\bm{C} | \theta, \bm{X}_1)\\
&= \prod_{i=1}^{N} \Bigg[  P(W_{i2}^{obs} | Z_{i2}, C_i) \cdot  P(W_{i1}^{obs} | Z_{i1}, C_i) \cdot  P(C_i | \theta, X_{i1}) \Bigg]
\end{align*}
then
\begin{align*}
P(\bm{C}| \theta, \bm{X}_1, \bm{Z}_1, \bm{Z}_2, \bm{W}_1^{obs}, \bm{W}_2^{obs}) &= \frac{\prod_{i=1}^{N} \Bigg[  P(W_{i2}^{obs} | Z_{i2}, C_i) \cdot  P(W_{i1}^{obs} | Z_{i1}, C_i) \cdot  P(C_i | \theta, X_{i1}) \Bigg]}{ \prod_{i=1}^{N} \sum_{C_i} \Bigg[  P(W_{i2}^{obs} | Z_{i2}, C_i) \cdot  P(W_{i1}^{obs} | Z_{i1}, C_i) \cdot  P(C_i | \theta, X_{i1}) \Bigg]}\\
=& \prod_{i=1}^{N} \frac{   P(W_{i2}^{obs} | Z_{i2}, C_i) \cdot  P(W_{i1}^{obs} | Z_{i1}, C_i) \cdot  P(C_i | \theta, X_{i1})}{ \sum_{C_i'} \Bigg[  P(W_{i2}^{obs} | Z_{i2}, C_i') \cdot  P(W_{i1}^{obs} | Z_{i1}, C_i') \cdot  P(C_i' | \theta, X_{i1}) \Bigg]}
\end{align*}

Therefore, compliance type follows a multinomial distribution with
\begin{equation}\label{secondstep}
P(C_i| \theta, X_{i1}, Z_{i1}, Z_{i2}, W_1^{obs}, W_2^{obs}) = \frac{   P(W_{i2}^{obs} | Z_{i2}, C_i) \cdot  P(W_{i1}^{obs} | Z_{i1}, C_i) \cdot  P(C_i | \theta, X_{i1}) }{  \sum_{C_i'} \Bigg[  P(W_{i2}^{obs} | Z_{i2}, C_i') \cdot  P(W_{i1}^{obs} | Z_{i1}, C_i') \cdot  P(C_i' | \theta, X_{i1}) \Bigg]}
\end{equation}


For step 3, based on the assumptions that given $\bm{C}$ and $\theta$, missingness does not depend on observed data, and past time points will not depend on the future time points, we can break it down into two parts:
\begin{itemize}
	\item Sample $\bm{X}_2^{mis} \sim P(\bm{X}_2^{mis} | \bm{C}, \theta, \bm{X}_1, \bm{Z}_1, \bm{W}_1)$
	\item Sample $\bm{Y}^{mis} \sim P(\bm{Y}^{mis} | \bm{C}, \theta, \bm{X}_2^{obs}, \bm{X}_2^{mis}, \bm{X}_1, \bm{Z}_1, \bm{Z}_2, \bm{W}_1, \bm{W}_2)$
\end{itemize}

\section{Conclusion}
This paper outlines the Bayesian causal inference framework in sequentially randomized experiments with non-compliant issues. We derived the posterior predictive distribution of the missing data and the sampling algorithm. For future work, we will use simulation studies and real data to demonstrate the validity of the proposed framework.

\bibliographystyle{apalike}
\bibliography{thesis}
\end{document}